# DIMENSIONAL ANALYSIS IN ECONOMICS:
# A STUDY OF THE NEOCLASSICAL ECONOMIC GROWTH MODEL

**Miguel Alvarez Texocotitla** *

**M. David Alvarez Hernández****

**Shaní Alvarez Hernández*****

**Abstract**

The fundamental purpose of the present research article is to introduce the basic principles of Dimensional Analysis in the context of the neoclassical economic theory, in order to apply such principles to the fundamental relations that underlay most models of economic growth. In particular, basic instruments from Dimensional Analysis are used to evaluate the analytical consistency of the Neoclassical economic growth model. The analysis shows that an adjustment to the model is required in such a way that the principle of dimensional homogeneity is satisfied.

---

\* Universidad Autónoma Metropolitana-Iztapalapa, Department of Economics. San Rafael Atlixco No. 186, Col. Vicentina, Iztapalapa, C.P. 09340, Ciudad de México, México. Email: atm@xanum.uam.mx.

\*\* Universidad Autónoma Metropolitana-Iztapalapa, Department of Physics. San Rafael Atlixco No. 186, Col. Vicentina, Iztapalapa, C.P. 09340, Ciudad de México, México. Email: mdalvarezh@gmail.com.

\*\* Universidad Autónoma Metropolitana-Iztapalapa, Department of Mathematics. San Rafael Atlixco No. 186, Col. Vicentina, Iztapalapa, C.P. 09340, Ciudad de México, México. Email: shanieneida@gmail.com.

**Introduction**

To understand the functioning of an economic system, relationships must be established between the fundamental variables of the system which are susceptible to be analysed by using mathematical models. However, economic modelling has analytical requirements that must be met, more so if the model intends to have a correspondence with reality and is not a mere mathematical abstraction. The requirements we will discuss in this article are within the field of Dimensional Analysis.

Dimensional Analysis studies the properties of observable quantities with dimensions and the properties of mathematical relationships that incorporate them (Sonin, 2001). This analysis is applied in the natural sciences; its principles (dimension, homogeneity, measurement and unity) are key in the formation of scientific thought since they are part of the basic principles of science. Compliance with the principles of Dimensional Analysis, and in particular the principle of dimensional homogeneity, is a basic prerequisite for proper mathematical modelling, as it allows to verify ex post the dimensional consistency of the mathematical relations and points out mathematical restrictions to the relations that are proposed between the variables.

However, in some disciplines as it is the case with Economics, the concept of dimension and their respective principles are practically unknown. Very few researches have emphasized the implications of Dimensional Analysis in the economic discipline; among them are Grudzewski and Roslanowska (2013), whose textbook provides an extensive work of the use and application of dimensional analysis in economic modelling; Barnett II (2004) analysed the dimensional consistency of production functions, and in particular the Cobb-Douglas function; Shone (2002) provided a brief but successful introduction to the use of dimensions in economic models; Okishio (1982) presented some application of dimensional analysis in the labour theory of value, and De Jong (1972), who was probably the first who pointed out the importance of dimensional congruence and its lack thereof in economic models. This scarcity of studies is one of the reasons that motivates us to reconsider this important area of mathematical modelling, since Dimensional Analysis can offer new insights into the way in which economic models are built, and in some cases, it can amend the errors and shortcomings that these models may contain.

On the other hand, research regarding dimensions represents a fundamental critique towards the methodology followed in economic modelling. Notwithstanding economic models that may be sustained by a solid mathematical theory, it may also be that it is not contemplated in their construction the restrictions that exist to model phenomena of various kinds. This situation would make necessary a thorough revision of the economic models from the perspective of Dimensional Analysis, in particular, of models of economic growth.

In this context, the present study's primary objective is to display the basic principles of Dimensional Analysis in the framework of economic theory, with the purpose to apply these principles to the fundamental relationships that underlie the models of economic growth. Specifically, dimensional analysis tools are used to evaluate the analytical consistency of the Neoclassical Growth model.[1] This analysis suggests the need for a correction in the model so that it complies with the principle of dimensional homogeneity.

At the end of this paper, some thoughts are offered about the current situation of economic modelling, the relevance of Dimensional Analysis in Economics, and the possible lines of research on this issue.

**1. Dimensional Analysis**

Scientific theories are logical-abstract structures that seek to explain and predict either natural, biological or social phenomena. The construction of any scientific theory begins with the observation and description of events of interest to infer (based on these observations and descriptions) laws, patterns and relationships that represent the phenomenon of interest in the most general and simplest way possible (Einstein, 1933). To accomplish this purpose, science uses mathematical language. We could assert that as long as a theory is not formulated with precision in the language of mathematics, it is not possible to assess its relevance and its ability to predict.

---

[1] The choice of this model in particular can be justified insofar as it constitutes the basic structure of numerous models of economic growth that have subsequently emerged.

However, the use of mathematical language has restrictions since its use entails the compliance of certain rules. These rules are related to the properties of the quantities that are included in a quantitative analysis.[2]

On the other hand, the use of mathematical tools is enriched with an axiom, which is derived from a very simple but fundamental idea: the mathematical relationships obtained from a scientific theory must relate phenomena of a similar nature, such that a causal relationship can be established. Using the jargon of the Dimensional Analysis, it can be said that the mathematical relations in any theory must be equidimensional.

The acceptance of this axiom implies the adoption of a validation criteria. A mathematical relationship is valid in the context of the theory, if on both sides of the equality (thinking in an algebraic equation, for instance) similar or equal terms are found. However, to say only intuitively that two observables are similar or equal is insufficient; how to define what is similar or what is different? It is in this context that the concept of dimension is introduced, which allows to manage the axiom of similarity using a more grounded mathematical formalism (White, 2011).

The axiom of similarity, or the principle of Dimensional homogeneity (as it is known in the literature), essentially expresses the following: If a mathematical relationship accurately represents a proper relationship between different observables in a certain theory, then this relationship is homogeneous, i.e. each of the additive terms of such relationship must be equidimensional.

**1.1 Basic aspects of Dimensional Analysis.**

The presentation of the basic principles of dimensional analysis requires to define some fundamental concepts. To start, it is necessary to explain what is meant by observable.[3]

---

[2] Here, quantity is meant as the description of any sensory perception or tangible property. For example, the notion of space (a physical property) is detectable through multiple quantities, either as distances, areas or volumes.

[3] This section is not intended to give a complete and exhaustive explanation of dimensional analysis. For a more detailed description, see the references.

**Definition 1.** *The quantities that are possible to quantify and include in a consistent theory (i.e. are describable by means of the mathematical language) are called observables.*[4]

Each observable represents different properties; consequently, to determine the possible interrelations between such observables it is necessary to establish some type of structure. To be able to compare two different observables, both must basically have defined an operation of causality or equivalence.

**Definition 2.** *We denote by $O$ the set of all possible observables that may exist within a theory:*

$$O \equiv \{A, B, C, \ldots\}$$

*Such set has the following structure:*

i) *An operation of causality ($=, \neq$), that allows to determine a relationship of order (equality or inequality) between two observables, e.g. $A = B$ or $A \neq B$.*

It should be emphasized that the causality operation has its justification in real experience, since the action of comparing two observables is closely related to the possibility of manipulation of objects or events that allow to perform any comparison or measurement that determines the cause/effect (Sonin, 2001). For example, there is no way to compare one cardinal observable (which represents the property of quantity) with a time observable (which represents the property of time), because in principle it is not possible to define a relation of equivalence that provides a result with real and physical sense.

Apart from the operation of causality, other mathematical operations can be defined for the observables. However, it is necessary to define the concept of dimension for this (Bunge, 1971).

**Definition 3**. *Dimension is the qualitative description of some physical property, which can be shared it by different types of observables. Let D be the set of all possible dimensions within a theory, associated with $O$:*

---

[4] Balaguer (2013), Bunge (1971), Carlson (1979), Fröhlich (2010), Sonin (2001) and White (2011) use the physical observable denomination. However, in this work we leave the physical adjective to emphasize that the rules of Dimensional Analysis equally apply to the economic observables.

$$D \equiv \{A, B, C, \ldots\}$$

*This set must have the structure of a commutative group, which is determined by:*

i)      *A binary operation* $(*)$,      *e.g.* $A * B = C$

ii)      *A neutral element* $(1_D)$,      *e.g.* $A * 1_D = A$

iii)      *An inverse element* $(A^{-1})$,      *e.g.* $A * A^{-1} = 1_D$

*And it must meet the following conditions:*

iv)      *The operation is associative,*      *e.g.* $A * (B * C) = (A * B) * C$

v)      *The operation is commutative,*      *e.g.* $A * B = B * A$

***Definition 4.*** *The allocation of dimensions to the observables is made through an operator* **[ ]: O→D**, *which is called the dimension function. This function assigns to each element of* **O** *a corresponding element in D, in such a way that*

i)      *If* **A, B ∈ O,** $A, B \in D$, *and* $\alpha, \beta$ *are real numbers, then* $[\alpha \mathbf{A}^\alpha \beta \mathbf{B}^\beta] = \alpha[\mathbf{A}^\alpha]\beta[\mathbf{B}^\beta] = \alpha A^\alpha \beta B^\beta$.

ii)      *If* $\alpha \in \mathbb{R}$, *then* $[\alpha] = 1_D$

The use of definition 4 can be exemplified. Consider an interest rate $\rho$, and the dimension associated to such observable is $T^{-1}$; then the result of applying the dimension function would be $[\rho] = T^{-1}$. If there was another object that did not belong to the set **O**, for instance any numeric value $r$, then the dimension function returns the neutral or non-dimensional element, $[r] = 1_D$. Like dimensions, the dimension function has the same algebra associated. The previous definitions offer a more formal definition of observable.

***Definition 5.*** *An observable* **A ∈ O** *is defined as the product of a numerical value* $\alpha \in \mathbb{R}$ *and a dimension* $A \in D$**:**

$$\mathbf{A} \equiv \alpha * [A]$$

Likewise, with the above definitions the concept of equidimensionality and the condition of equivalence between observables is shown.

**Definition 6.** *Let $A, B \in O$, and $C \in D$. Then, $A$ and $B$ are equidimensional if and only if they have the same dimension, i.e.:*

$$[A] = C = [B]$$

**Definition 7.** *Let $A, B \in O$ be two observables, α, β numerical values, and $A, B \in D$ the dimensions associated with $A$, $B$ respectively. Then $A$ and $B$ are equivalent (equal) if and only if they have the same numerical value and the same dimension. That is:*

$$A = B \text{ if and only if } \alpha = \beta \text{ and } [A] = [B]$$

*Consequently:*

i) *The observables have a binary operation ($*/$) of product/division defined.*

ii) *Only the equidimensional observables have a binary operation of addition/subtraction ($\pm$) defined.*

iii) *The arguments of transcendental functions, e.g. exponentials and logarithms, should be strictly dimensionless terms.*

iv) *The ratio of two equidimensional observables gives as a result a dimensionless observable.*

**1.2 The Principle of Dimensional Homogeneity**

Scientific theories typically involve two types of equations: mathematical and observable (Sonin, 2001). Mathematical equations contain only numeric values or other mathematical entities that do not have any innate real meaning, which does not imply that they cannot be explored and studied to find relationships and properties arising from such equations.

Conversely, observable equations arise primarily from observation and experimental data. The importance of these equations lies on the fact that they relate in a balanced manner a set of observables against another set of observables through the sign of mathematical equality (Baiocchi, 2012).

Since the observable equations contain observable quantities, which necessarily carry a real meaning, it is essential to be careful that the equation that is being proposed is properly balanced. Consequently, the use of dimensions essentially provides the criterion of balance in the equations. The following definition expresses the dimensional equality (balance) of an equation.

**Definition 8.** *An equation is equidimensional if and only if each of the additive observables that compose it is equidimensional.*

In other words, an observable equation is valid only if it is an equidimensional equation, that is, if all the terms involved in the equation are equidimensional. This simple idea marks the most fundamental principle of Dimensional Analysis, and such is its importance that the idea rises to the level of axiom (Sonin, 2001).

**Axiom or Principle of Dimensional Homogeneity.**

*An equation that contains observables is valid and has a real meaning in the context of the theory, if and only if, each of the observables involved in the equation are equidimensional amongst themselves.*

The origin of this axiom predates the formalization of Dimensional Analysis, since to calculate a result from any equation it is imperative to have congruency between the result and the terms or observables which originated such result, i.e., a cause/effect relationship can only be set between phenomena of the same nature (Macagno, 1971). The following theorem is derived as a result of the axiom.

**Theorem.** *The numerical validity of an equation that involves observables exists if and only if the terms on both sides of the equality are equidimensional.*

Consider the following example. A theory considers that the only relevant observables are the number of available machines *X*, the total number of workers *P*, the number of inactive workers $P_I$ and the number of produced machines *Y*. It is proposed that the number of produced machines is given by the following equation:

$$Y = X + (P - P_I)$$

By the principle of dimensional homogeneity, we can only calculate a valid result for *Y* if and only if $X, P, P_I, Y$ are equidimensional. The term *X* is equidimensional with *Y*, since both represent two quantities of the same observable, thus it is possible to establish a relationship of equality between both terms. Terms *P* and $P_I$ are equidimensional among themselves, and therefore the subtraction of the two has physical sense. However, *P* and $P_I$ are not equidimensional with respect to *X* or *Y*, thus it makes no sense to establish a relationship of addition between those observables, as expressed in the equation. Seen from the dimensional point of view, equation *Y* lacks all sense and real meaning with the observables currently involved, given that there is no way to establish the meaning of subtracting or adding the number of machines with the number of workers. The only way to validate equation *Y* is by incorporating different observables or additional terms in order to homogenize the equation's dimensions.

## 2. Dimensional Analysis in Economics

This section applies the principles of Dimensional Analysis to economic theory and analyzes some fundamental economic relationships that underlie most of the models of economic growth. It is shown that these relationships suffer from inconsistencies from the point of view of Dimensional Analysis, which are possible to correct at the cost of facing new analytical implications.

Dimensional Analysis has its roots in the first primitive notions of pre-scientific thought that several cultures have constructed to describe the physical world and explain its functioning in quantitative terms. For example, its origin in western thought can be traced back from the geometric formulations performed by the Greek mathematicians of antiquity. However, its modern foundation and form began to be established three centuries ago, and it was mainly in the twentieth century that its use was adopted extensively in the natural sciences.

As far as it is known, the first person who wrote about the problem of units and dimensions (essentially focused on physics' models and theories) was Leonard Euler in 1765. Later on, the subject was again studied by Joseph Fourier in his book *The Analytic Theory of Heat*, published in 1822, were he remarked what is now known as the principle of dimensional homogeneity, and even developed some of the common rules that are currently used in Dimensional Analysis. Afterwards, there were no significant advances in the area until the publication of the book the *Theory of Sound*, by Lord Rayleigh in 1877, which proposed a method very similar to the modern method of dimensions, where gave numerous examples of how to use the methods of Dimensional Analysis. Notwithstanding the aforementioned contributions, the most notorious one that established the method and the current rules of Dimensional Analysis was that of E. Buckingham, who in 1914 established the most used formalization of Dimensional Analysis (Macagno, 1971).

**2.1 Stocks, Flows and Economic Dimensions**

Although the concepts of Dimensional Analysis are practically unknown in Economics, some intuition has permeated through the concepts of stock and flow. Shone (2002) points to the distinction between two main types of economic observables, the stock-type observables and the flow-type observables.

The stock observables refer to the value of a certain economic quantity at a certain date (i.e. at an exact point in time). Suppose that our economic observable is the amount of money ($Ms$) in the economy. This observable would have a defined value for each moment of time, for example as of December 31, and will have another defined value ($M's$) for any other instant in a previous or later time. That is, the dimensional characteristic of a stock observable is independent of any time interval.

On the contrary, a flow observable refers to the total or average value of a certain economic quantity in a time interval. For example, if we consider the demand for goods ($D$) in a given period of time, like in a year, such observable is of flow-type, since it is an observable that is distributed over time, and thus its value is defined in that same interval. In other words, the flow observable has a direct dimensional dependency over time.

If we assume that the stock and flow observables are represented by continuous and differentiable functions, the relationship between a stock and a flow can be represented mathematically as follows: [5]

$$\frac{dQ(t)}{dt} = F(t) \qquad (1)$$

Where $Q(t)$ is the function associated with a stock-type observable and $F(t)$ is the function associated with the flow-type observable. Therefore, since the derivative is itself a mathematical operator that depends on time, any stock observable can be transformed by the derivative operator into a flow observable. [6]

Another way to see the difference between stock and flow is to think of $F(t)$ as the equivalent of a change rate (i.e. a velocity) of a given quantity or stock, in this case $Q(t)$. Consequently, given their difference, the only way in which both observables could be equal, i.e. equidimensional, would be to use another term that homogenizes the dimensions of the relation. In this case, the term that homogenizes the relation is the derivative operator, because this operator transforms $Q(t)$ from a stock to a flow $F(t)$.

---

[5] This definition can also be expressed in its discrete time equivalent.
[6] Note that even though the variables are functions of time, this does not imply that they incorporate a time-dependent dimension. Only when a time-dependent observable is explicitly incorporated, such as the derivative operator, the dimensions of the variable incorporate a time-dependent dimension (T).

Therefore, a stock and a flow are not equidimensional, that is, they cannot be compared, equalized, added or subtracted, since these operations are not defined to operate between non-equidimensional observables. This difference in the dimensions of stocks and flows imposes restrictions that must be taken into consideration, especially when combining observables of different natures.

Once the concepts of stock and flow have been discussed, it is appropriate to consider what kind of dimensions could be used in Economics. Outside the field of Physics, the choice of the fundamental dimensions is not so straightforward, as it depends very much on the area of application and the theory used. For example, in the case of neoclassical economic theory, it seems to be sufficient to have as fundamental dimensions the time ($T$), the monetary value ($M$), the quantity of elements ($Q$) and the utility ($U$).[7] This set of dimensions presents a good starting point to incorporate the question of dimensions in the economic growth models, since there seems to be no additional fundamental economic characteristics than those represented by these four dimensions.[8] For this reason, they will be used throughout this article.

**2.2 The Dimensional Analysis in the models of economic growth.**

In this section we analyse two identities that are fundamental in the models of economic growth: the identity of income of the representative household and the identity of profit of the representative firm. Both identities derive the relationships that determine the dynamics of production and the accumulation of physical capital.

Consider the case of a model of economic growth for a closed economy that contains only two economic agents, a household and a firm.[9] Both exchange in the market a single good of consumption/production, which is denoted as physical capital. This unique representative good is used simultaneously as a production input and as a consumer commodity. [10]

---

[7] This set of dimensions is proposed in Shone (2002).
[8] However, the justification as of why that set is adequate and sufficient is an issue that is yet to be determined.
[9] Both can be considered as aggregates or representatives of all other agents of the system, thus they also receive the names of representative household and representative firm.
[10] As in the case of the representative household and firm, this good can be considered as the aggregate of all other goods in the economy.

In addition to the existence of a single good, it is assumed that there is another economic factor that only affects the productive process (that is, it only works as a production input). This factor is called labour. The combination of both factors (physical capital and labour) takes place in the productive process and leads to the production of more physical capital, which may or may not be used again in the productive process.

Following the conventional assumptions of the models of economic growth, it is considered that the structure of the market where both agents participate is of perfect competition. The household and the firm act as agents that take their decisions regardless of market prices associated with production inputs and the representative good.

On the other hand, it is assumed that the household is the owner of the factors of production and its income $Y_H$, for the rent of these inputs to the firm, will be given by:

$$Y_H(t) = w_K(t)K_H(t) + w_L(t)L_H(t) \qquad (2)$$

Where $w_K$ and $w_L$ denote the rental prices of the physical capital $K_H$ and labour $L_H$. (Acemoglu, 2009, pp. 32).

Considering the dimensions of each additive term, the dimensional congruence of Eq. (2) is now analysed. The physical capital represents the amount of goods that exist in an instant of time, i.e. a stock observable. The labour, unlike the physical capital, represents a flow observable, as it is measured in terms of working units used per time interval. [11] Consequently, the dimensions associated with $K_H$ and $L_H$ are:

$$[K_H] = Q_K, \qquad [L_H] = \frac{Q_L}{T}$$

With $Q_K$ symbolizing the dimension of physical capital quantity, $Q_L$ the dimension of labour quantity and $T$ the time dimension.

The income $Y_H$ is expressed in terms of monetary units received in a time interval, e.g. the income reported by an individual for the rent of a service may be expressed in terms of dollars/per month. Therefore, $[Y_H] = \frac{M}{T}$. However, in the context of the model that develops, there is no Central Bank and therefore there is not a monetary unit. That is, we are considering a non-monetary economy of exchange of goods for goods. Consequently, income is a flow-type economic observable of physical capital goods, i.e. the replacement $M \to Q_K$ is made and as such its dimensions are expressed as:

$$[Y_H] = \frac{Q_K}{T}$$

---

[11] See Pindyck and Rubinfeld (2009, pp. 219).

The prices, $w_K$ and $w_L$ represent how many monetary units are equivalent to a physical capital unit or a unit of work, but in the context of a non-monetary economy, prices only represent how many units of physical capital are equivalent to one unit of work. Thus, their dimensions are the following: [12]

$$[w_K] = \frac{M}{Q_K} = \frac{Q_K}{Q_K} = 1_D, \quad [w_L] = \frac{M}{Q_L} = \frac{Q_K}{Q_L}$$

Consequently, rewriting the Eq. (2) in terms of the dimensions of each variable:

$$[Y_H] = [w_K][K_H] + [w_L][L_H] = (1_D)(Q_K) + \left(\frac{Q_K}{Q_L}\right)\left(\frac{Q_L}{T}\right)$$

$$\left(\frac{Q_K}{T}\right) \neq (Q_K) + \left(\frac{Q_K}{T}\right)$$

Therefore, comparing the dimensions of each additive term, it is found that:

$$[Y_H] \neq [w_K][K_H], \qquad [Y_H] = [w_L][L_H] \tag{3}$$

And this shows the first major result of the dimensional analysis of the income identity Eq. (2): the relationship that describes household income is not dimensionally congruent, as it is not satisfied that on both sides of the equality the same dimensions are found. [13] Where does this error come from?

---

[12] Notice that under the assumption of a non-monetary economy, the price of physical capital becomes an adimensional quantity.
[13] See Section 1.

The dimensional incongruence is generated by the dissimilar dimensional nature of physical capital and labour. It is assumed in Eq. (2) that both factors, physical capital and labour, can be added to each other simply because both factors are multiplied by their respective prices, however, both factors have a different nature, as one describes a stock while the other factor describes a flow.

The firm side is now examined. It is considered theoretically that the firm acts according to the maximization of its profit. Likewise, it is assumed that the productive process depends only on the inputs of production (physical capital and labour) and that this process can be described by a function of production $F(K_F; L_F)$:

$$Y_F = F(K_F; L_F) \qquad (4)$$

Therefore, the benefit of the firm will be given by:

$$\pi = F(K_F; L_F) - (w_K(t)K_F(t) + w_L(t)L_F(t)) \qquad (5)$$

Where $\pi$ denotes the benefit of the firm reported in physical capital units per time interval, $Y_F$ denotes the production carried out by the firm, expressed similarly in units of physical capital per period of time, $w_K$ and $w_L$ are the prices of the physical capital and the labour, $K_F$ represents the amount of physical capital used by the firm, and $L_F$ the number of labour units used in the production time interval.

As before, Eq (5) is rewritten in terms of the dimensions of each term:

$$[\pi] = [Y_F] = \frac{Q_K}{T}, \qquad [F(K_F, L_F)] = \frac{Q_K}{T}, \qquad [K_F] = Q_K, \qquad [L_F] = \frac{Q_L}{T}$$

$$[w_K] = 1_D, \qquad [w_L] = \frac{Q_K}{Q_L}$$

Thus,

$$[\pi] = [F(K_F; L_F) - (w_K K_F + w_L L_F)]$$

$$\left(\frac{Q_K}{T}\right) = \left(\left(\frac{Q_K}{T}\right) - (1_D)(Q_K) + \left(\frac{Q_K}{Q_L}\right)\left(\frac{Q_L}{T}\right)\right) = \left(\frac{Q_K}{T}\right) - (Q_K) + \left(\frac{Q_K}{T}\right)$$

Consequently, an inconsistency is found in the dimensions of Eq. (5):

$$[\pi] \neq [w_K K_F] \tag{6}$$

Like the inconsistency of Eq. (2), it derives from the fact that two terms of different nature are compared (added in this case); it compares a stock quantity (physical capital) with a flow quantity (labour).

How can this dimensional inconsistency be corrected? There are two alternatives. The first is to consider a different interpretation of physical capital so that it assumes the dimensions corresponding to a flow $\left([K] = Q_K/T\right)$, so that the term $w_K K$ acquires the correct dimensions. However, it is difficult to justify a different interpretation of $K$ when it is a component of the income identity.

The second option is to consider a different interpretation of the term $w_K$. Suppose that instead of $w_K$, a return rate (yield) is introduced, i.e. $(w_K \rightarrow r_K)$, which would represent how much physical capital is given by a physical capital unit in each time interval.

On the other hand, it should be taken into account that the physical capital can be incorporated or be out of the productive inputs market. That is, new capital is incorporated into the market, but all the same, loses of capital occur when it leaves the market because of wear, destruction, breakdown, or turning obsolete. Under these considerations it is also justified to introduce a capital depreciation rate, which will be considered a constant for simplicity, i.e. $\left(\dot{K}(t)/K(t)\right) = -\delta$ (Barro & Sala-i-Martin, 2009, pp. 32). Therefore, if these changes enter in Eq. (2) and in Eq. (5), considering that the dimensions of $r_K$ and $\delta$ are:

$$[r_K] = \frac{Q_K}{Q_K T} = \frac{1}{T} = [\delta]$$

it is found that the dimensional inconsistency of Eq. (2) and Eq. (5) is resolved, as can be seen below.[14]

---

[14] The sub-indexes of K and L have been eliminated, given that we are considering that the market is in equilibrium. In other words, the quantities of capital and labour that the firm requires are exactly matched

$$Y = r_K(t)K(t) + w_L(t)L(t) \tag{7}$$

$$[Y] = \frac{Q_K}{T} = [r_K K] = [w_L L]$$

$$\pi = F(K; L) - \big((r_K(t) + \delta)K(t) + w_L(t)L(t)\big) \tag{8}$$

$$[\pi] = \frac{Q_K}{T} = [F(K; L)] = [r_K K + w_L L]$$

The introduction of $r_K$ must not seem strange; Barro & Sala-i-Martin (2009) emphasize this term by denominating it correctly as a yield rate. It is important not to confuse the concepts of income price and yield rate, since both represent two observables of a different nature. Prices only convert stocks into stocks, while rates transform stocks into flows.

Also, to preserve dimensional symmetry in Eq. (7) and Eq. (8), the labour factor $L$ can be reformulated in terms of the variation of a new variable, $P$, of stock-type:[15]

$$L \equiv \frac{dP(t)}{dt} = \dot{P}(t) = nP(t) \tag{9}$$

Which is assumed to grow exponentially and therefore at a constant growth rate of $\big(\dot{P}(t)/P(t)\big) = n$.[16]

The variable $P$ is called population stock and must be interpreted as the number of inhabitants or economic agents within the representative household. Thus, the dimension associated with this variable is the quantity of population $[P] = Q_P$. This reformulation of the labour factor is congruent from a dimensional point of view. The introduction of the derivative operator does not contravene the original dimensions of $L$:

$$[L] = \frac{Q_P}{T} = \left[\frac{dP(t)}{dt}\right] \tag{10}$$

---

with the quantities of capital and labour that the representative household rents ($L_F = L_H = L$, $K_F = K_H = K$).

[15] This reformulation is grounded in the hypothesis that when the market is in equilibrium, all the labour available is used. Thus, each member of the work force gives exactly one unit of labour, independently of the wage associated to such factor, i.e. $Q_P = Q_L$.

[16] In order to simplify the notation, the time dependency of the variables will not be pointed out explicitly. When differentiation with respect to time is performed, Newton's dot notation will be used. When differentiation with respect to other variables is performed, Leibniz's notation will be used.

Also, the introduction of the variable $P$ allows to transform the original variables correctly into per capita variables.

$$k(t) \equiv \frac{K(t)}{P(t)}, \quad n \equiv \frac{L(t)}{P(t)}, \quad f(k;n) \equiv \frac{F(K;L)}{P(t)}, \quad y(t) \equiv \frac{Y(t)}{P(t)}$$

It should be noted that when Eq. (9) is entered in the production function, the function not only depends on the physical capital per capita, but also depends on the growth rate of $P(t)$. This fact comes from the homogeneity property of the production function.[17]

To determine exactly the effect of the modifications made in equations Eq. (7), Eq. (8), and the introduction of $P(t)$ in the production function, the function $(k; n)$ is explicitly calculated. Starting with Eq. (7), the equation is rewritten in per capita variables and is derived with respect to time:

$$\dot{y}(t) = \dot{r}(t)k(t) + r(t)\dot{k}(t) + \dot{w}(t)n$$

Developing the expression,

$$\frac{\dot{y}(t)}{y(t)} = \frac{r(t)k(t)}{y(t)}\left(\frac{\dot{r}(t)}{r(t)}\right) + \frac{r(t)k(t)}{y(t)}\left(\frac{\dot{k}(t)}{k(t)}\right) + \frac{w(t)n}{y(t)}\left(\frac{\dot{w}(t)}{w(t)}\right)$$

$$\frac{\dot{y}}{y} = \beta\left(\frac{\dot{r}}{r}\right) + \beta\left(\frac{\dot{k}}{k}\right) + (1-\beta)\left(\frac{\dot{w}}{w}\right) \tag{11}$$

where $\beta(t) \equiv \frac{r(t)k(t)}{y(t)}$ has been defined as the participation of capital in production, and $(1 - \beta(t)) \equiv \frac{w(t)n}{y(t)}$ as the participation of labour.

If the hypothesis that the shareholdings of capital and labour remain constant is taken into account, i.e. $\beta(t) = \alpha$, and the yield of capital and the price of labor remain unchanged, i.e., $\dot{r} = \dot{w} = 0$, then Eq. (11) can be integrated directly. The result is the Cobb-Douglas function:

---

[17] A requisite that any production function must comply is the constant returns to scale hypothesis. In other words, the function must be a homogeneous first order function in each one of its arguments. See Acemoglu (2009, p. 29).

$$y(t) = f(k;n) = a_0 k^\alpha, \qquad Y(t) = F(K;L) = a_0 K^\alpha \left(\frac{L}{n}\right)^{1-\alpha} \qquad (12)$$

With $a_0$ a constant of integration. This calculation validates the modifications made to Eq. (7) and Eq. (8), since the Cobb-Douglas production function has been explicitly obtained. The new representation of labour as a variation of the population stock, seen in Eq. (9), is also justified, as well as its incorporation into the production function.

## 3. Dimensional Analysis of the Neoclassical Economic Growth Model

This section analyses the dimensional consistency of the Neoclassical model and incorporates the corrections proposed in the second section.

The basic version of the Neoclassical model is built in the context of a closed economy, in which there exists a single market where a single good is produced and consumed; that is to say, this good can be transformed either into a production input (e.g. physical capital) or into a consumer good. Likewise, it is assumed that the principle of aggregation is valid (existence of a representative agent) and therefore the general problem of optimization of all the agents that participate in the market can be reduced to solving a single problem of optimization.[18]

The main difference of the Neoclassical model with respect to the first models of growth, such as the Solow-Swan model, is that the Neoclassical model explicitly considers the consumption of the representative house, and in this way endogenizes the savings, which was considered as an exogenous variable in the previous models.

In the Solow-Swan model, it is assumed that the investment (in the context of a closed economy and perfect competition market) is equal to the savings function, and that the savings function is equal in turn to a fixed (exogenous) proportion of the income (Solow, 1956). In the Neoclassical model, this assumption is no longer considered. Household savings are now explicitly described in terms of the income $Y(t)$ and consumption $C(t)$:

---

[18] For a more detailed explanation of the Neoclassical model, it is recommended to review Cass (1965), Samuelson (1970) and Acemoglu (2009).

$$I(t) = S(t) = Y(t) - C(t) = r_K(t)K(t) + w_L(t)L(t) - C(t) \qquad (13)$$

Thus, the equation that describes the accumulation of capital of the representative household is given by:

$$\frac{dK(t)}{dt} = I(t) = r_K(t)K(t) + w_L(t)L(t) - C(t) \qquad (14)$$

To reduce the number of variables involved in Eq. (14) and obtain a differential equation in one variable, per capita variables are introduced. Thus: [19]

$$\dot{k}(t) = r_K(t)k(t) + nw_L(t) - c(t) - nk(t) \qquad (15)$$

On the other hand, the condition of maximizing the profit of the firm implies that in the market equilibrium, the production function must satisfy the following conditions (Acemoglu, 2009, pp. 33):

$$\frac{\partial F}{\partial K} = r_K(t) + \delta, \qquad \frac{\partial F}{\partial L} = w_L(t)$$

These conditions can be rewritten in terms of per capita variables:

$$f'(k;n) = r_K(t) + \delta, \qquad f(k;n) - kf'(k;n) = nw_L(t)$$

Introducing these conditions into Eq. (15), a modified Solow-Swan equation is obtained:

$$\dot{k}(t) = f(k;n) + (n + \delta)k(t) - c(t) \qquad (16)$$

Equation. Eq. (16) is similar to the original equation shown in the literature,[20] except for the difference that the production function per capita presents, because now the function depends explicitly on the growth rate, $n$.

---

[19] Per capita consumption has been defined as: $c(t) \equiv \frac{C(t)}{P(t)}$
[20] See Acemoglu (2009), Barro and Sala-i-Martin (2009) or Solow (1956).

The goal of Eq. (16) is to describe the accumulation of physical capital per capita in function of two factors: the production function $f(k;n)$ and the consumption per capita $c(t)$.

To solve this equation and study the dynamics of physical capital, it is necessary to define the explicit form of the production function and propose a dynamic for the consumption. It is in this context that the model is interpreted as an optimal control problem.[21]

In the case of the Neoclassical model, what is optimized (maximized) is the welfare of society in a time interval (also called planning horizon). The fundamental hypothesis of the model is that the welfare of society at every moment of the planning horizon can be modelled through a utility function, which depends on the consumption level only. In this way, the model considers it permissible to assume the existence of a single agent (representative household) that represents all the economic decisions that are made.

This agent seeks to optimize its well-being, which is represented by a utility function $U(t)$ that depends exclusively on the consumption of the agent:[22]

$$u(t) = u(c(t)) \qquad (17)$$

However, the optimization is performed intertemporally, considering the future flow of the utility. Therefore, the agent optimizes the present value of its utility in an infinite time horizon $[0, \infty]$; that is, it maximizes the following expression:

$$u_P = \int_0^\infty u(c(t)) e^{-\rho t} dt \qquad (18)$$

The cost functional, Eq. (18), depends only on two variables, the state variable $k(t)$ and the control variable $c(t)$. The presence of the multiplier $e^{-\rho t}$ is due to the discount of the future value of the utility, which is made with a subjective discount ratio $\rho$ strictly greater

---

[21] In general, the optimal control problems seek to optimize a functional cost $J$, which depends on the variables of state $x_i(t)$, which describe the dynamics of the system, and the control variables $u_i(t)$. These variables can be controlled or chosen in such a way that they optimize the cost functional.

[22] The utility function has been defined in terms of per capita variables: $u(c(t)) \equiv \frac{U(C(t))}{P(t)}$

than zero $(\rho > 0)$; this condition ensures that the cost functional, Eq. (18), is convergent and it also guarantees the existence of solutions for the optimization problem (Acemoglu, 2009, pp. 252).

Conversely, the optimization problem is subject to satisfying the evolution of the state variable. In this case, as the variable of state is the physical capital per capita $k$, the optimization will be subject to satisfying the dynamic equation of the capital, i.e. the Solow-Swan modified equation, Eq. (16), (Cass, 1965).

In short, the Neoclassical model considers that the problem of economic growth can be treated as an optimal control problem by assuming that the representative agent seeks to maximize its present utility:

$$max \; u_P = max \left[ \int_0^\infty u(c(t)) e^{-\rho t} \, dt \right] \quad (19)$$

Subject to the restriction of capital accumulation, Eq. (16):

$$\dot{k}(t) = f(k; n) - (n + \delta) k(t) - c(t) \quad (20)$$

Once the bases of the Neoclassical model are specified, its dimensional analysis is carried out. Calculating first the dimensions of each term of Eq. (19):

$$[u_P] = (U) \left( \frac{1}{Q_P} \right) = \frac{U}{Q_P}$$

$$[u(c(t)) \, e^{-\rho t} dt] = [u(c(t))] [e^{-\rho t}] [dt] = \left( \frac{U}{Q_P T} \right) (1_D) (T) = \left( \frac{U}{Q_P} \right)$$

In this manner:

$$[u_P] = \frac{U}{Q_P} = [u(c(t)) e^{-\rho t} dt]$$

Therefore, the cost functional does not have any dimensional inconsistencies, since by integrating the utility flow $u(c(t))$ with respect to time, said term is transformed into a utility stock. Similarly, the discount term does not introduce any problems, given that the dimensions of the exponential's argument are nullified, and the term as a whole is dimensionless.

The calculation of the dimensions of each term of the constraint equation, Eq. (20), are continued:

$$[\dot{k}] = \left(\frac{Q_K}{Q_P}\right)\left(\frac{1}{T}\right) = \frac{Q_K}{Q_P T}, \quad [f(k;n)] = \left(\frac{1}{Q_P}\right)\left(\frac{Q_K}{T}\right) = \frac{Q_K}{Q_P T}$$

$$[(n+\delta)k] = \left(\frac{1}{T}\right)\left(\frac{Q_K}{Q_P}\right) = \frac{Q_K}{Q_P T}, \quad [c] = \left(\frac{1}{Q_P}\right)\left(\frac{Q_K}{T}\right) = \frac{Q_K}{Q_P T}$$

$$[\dot{k}(t)] = \left(\frac{Q_K}{Q_P T}\right) = [f(k;n) - (n+\delta)k(t) - c(t)] = \left(\frac{Q_K}{Q_P T}\right) - \left(\frac{Q_K}{Q_P T}\right) - \left(\frac{Q_K}{Q_P T}\right)$$

$$[\dot{k}(t)] = [f(k;n) - (n+\delta)k(t) - c(t)] \tag{21}$$

Calculating the dimensions of Eq. (20) shows that it is dimensionally congruent. It is important to remark that the dimensional consistency of Eq. (19) and Eq. (20) is guaranteed when the variables per capita with respect to the stock of population are defined.

What effect do these modifications have on the model? To identify the effects, the model must be resolved.

The solution to the optimal control problem is carried out using the methodology of Pontryagin's Maximum Principle.[23] This principle establishes a set of necessary conditions, such that the solution trajectories of the variables of state and control can be considered optimal. These conditions are expressed in terms of a function, which is known as the Pontryagin's Hamiltonian.

In the case of the Neoclassical model, there is only one state variable, which is the stock of physical capital per capita $k(t)$, a control variable, which is the consumption per capita $c(t)$, and a single restriction function, which is the equation of accumulation of capital, Eq. (20). Thus, the Hamiltonian associated with the model is the following:

$$H(k,c,p,t) = u(c)e^{-\rho t} + p(t)(f(k;n) - (n+\delta)k(t) - c(t)) \quad (22)$$

where $p(t)$ denotes the co-state variable. [24] Since the objective function of the model depends explicitly on the time due to the discount factor $e^{-\rho t}$, the model must be transformed into a problem of autonomous optimization, by using the definition of an autonomous Hamiltonian (also called of current time):

$$\hat{H} = He^{\rho t} \quad (23)$$

---

[23] For a more in-depth look at this methodology, it is recommended to review Acemoglu (2009, pp. 227-275) and Pedregal (2004, pp. 195-224).

Likewise, a co-state variable is defined in current time as:

$$\mu(t) = p(t)e^{\rho t} \quad (24)$$

Therefore, by rewriting Eq. (22):

$$\hat{H}(k,c,p,t) = u(c) + \mu(t)(f(k;n) - (n+\delta)k(t) - c(t))$$

For the trajectories of the state variables $k(t)$ and control $c(t)$ to be optimal and to satisfy the restriction equation, the Hamiltonian in current time must satisfy the conditions:[25]

$$\frac{\partial \hat{H}}{\partial c} = 0 = u'(c) - \mu(t) \quad (25)$$

$$\dot{k}(t) = \frac{\partial \hat{H}}{\partial \mu} = f(k;n) - (n+\delta)k(t) - c(t) \quad (26)$$

$$\dot{\mu}(t) = -\frac{\partial \hat{H}}{\partial k} + \rho\mu(t) = -\mu(t)\big(f'(k;n) - (n+\delta)\big) + \rho\mu(t) \quad (27)$$

---

[24] The coestate variable is interpreted as the shadow price of the capital stock. In other words, it is the value of a unit of capital in terms of utility. Thus, the dimensions of this variable are: $[p] = {^U}/{_{Q_K}}$.

[25] See Acemoglu (2009, p. 235).

Equations (25-27) form the system of differential equations whose solution will determine the optimal solutions of consumption and of physical capital per capita.

Before continuing with the solution of the model, the dimensions of the previous equations are analysed:

$$[\hat{H}] = \frac{U}{Q_P T} = [u(c)] = [\mu(t)][f(k;n)] = [\mu(t)][(n+\delta)k(t)] = [\mu(t)][c(t)]$$

$$\left[\frac{\partial \hat{H}}{\partial c}\right] = \left(\frac{U}{Q_K}\right) = [u'(c)] = [\mu(t)]$$

$$[\dot{k}(t)] = \left(\frac{Q_K}{Q_P T}\right) = \left[\frac{\partial \hat{H}}{\partial \mu}\right]$$

$$[\dot{\mu}(t)] = \left(\frac{U}{Q_K T}\right) = \left[\frac{\partial \hat{H}}{\partial k}\right] = [\rho \mu(t)]$$

The results of the dimensional analysis of the Hamiltonian and of the equations (25-27) allow us to indicate that these equations are dimensionally consistent. By solving the system of equations (25-27), these are rewritten in the following way:

$$u'(c) = \mu(t) \tag{28}$$

$$\dot{\mu}(t) = \mu(t)\big(\rho + n + \delta - f'(k;n)\big) \tag{29}$$

$$\dot{k}(t) = f(k;n) - (n+\delta)k(t) - c(t) \tag{30}$$

Then, with the purpose to obtain a system of only two differential equations, Eq. (28) is eliminated when it is introduced in the other two relations (29-30). Consequently, the dynamic equation for the consumption per capita is obtained:

$$\dot{c}(t) = \frac{c(t)}{\varepsilon_u(c)}\big(f'(k;n) - (\rho + n + \delta)\big) \tag{31}$$

where the term $\varepsilon_u$ represents the marginal elasticity of the utility function.[26] The Cobb-Douglas function is also explicitly introduced in Eq. (30) to fully express the dynamic equation of the capital:

$$\dot{k}(t) = a_0 k^\alpha - (n+\delta)k(t) - c(t) \tag{32}$$

Eq. (31) and Eq. (32) determine the optimal consumption trajectories and physical capital per capita trajectories for the optimal control problem of the modified Neoclassical model. The equidimensionality of the modified model is corroborated by the analysis of Eq. (31) and Eq. (32):

$$[\dot{k}(t)] = \left(\frac{Q_K}{Q_P T}\right) = [a_0 k^\alpha] = [(n+\delta)k(t)] = [c(t)]$$

$$[\dot{c}(t)] = \left(\frac{Q_K}{Q_P T^2}\right) = \left[\frac{c(t)}{\varepsilon_u(c)} f'(k;n)\right] = \left[\frac{c(t)}{\varepsilon_u(c)}(\rho + n + \delta)\right]$$

A thorough analysis of the dynamical properties of the model with the proposed correction is beyond the scope of this article. Nevertheless, it can be stated that the most appropriate way to analyse the optimal trajectories, Eq. (31) and Eq. (32), is through a phase space analysis, since such analysis allows the visualization of the dynamic behaviour of the trajectories in a global way, as well as the stability analysis of the equilibrium points of the system, which provide the fundamentals characteristics of the dynamics.[27]

The analysis developed throughout this last section shows that the Neoclassical model complies with the principle of dimensional homogeneity, as long as the modifications made to the income, Eq. (7), and benefit, Eq. (8), are accepted. Likewise, the introduction of per capita variables in terms of population stock is a necessary requirement, for the reason that otherwise the model would fail to comply with the principle of homogeneity by not properly relating the capital exchange rate per capita with the production function.

---

[26] It should be considered that the elasticity coefficients are dimensionless numbers, and in this case the adimensionality of the marginal elasticity of the utility is quickly verified with a little bit of algebra:
$[\varepsilon_u(c)] = \left(\frac{UT^2 Q_P^2}{Q_K^2}\right)\left(\frac{Q_K}{Q_P T}\right)\left(\frac{UTQ_P}{Q_K}\right)^{-1} = 1.$

[27] See Barro & Sala-i-Martin (2009) or Shone (2002).

The theoretical implications of the corrected Neoclassical model (equidimensional) have not yet been fully explored nor compared to the results of the original model. However, the implications of the equidimensional model are likely to be more relevant to understanding the dynamics of economic growth, as they would be freed of the dimensional inconsistency of the original model.

**Conclusions**

The economic theory has strived to have the same level of success in understanding and predicting economic phenomena as the success achieved by science in understanding and predicting natural phenomena. For this reason, the economic theory has emulated the analytical, statistical and mathematical methods of the natural sciences, restricting itself in a high degree to the formulation of mathematical models to explain the economic reality. That is to say, this theory tends to become a mathematical modelling exercise rather than a true scientific discipline of observation. In this way, although the economists claim that Economics is a science as objective and solid as the natural sciences, nowadays the discipline is built on theories, assumptions and mathematical models that are flimsy and highly questionable from an analytical point of view, as this research has tried to demonstrate.

By emphasizing the idealization of the technique and formal elegance of mathematics, the economic theory has become a sort of social mathematics that employs economic terms. This tendency towards formalism in Economics has had it consequences; formality and mathematical elegance have been overvalued, to the detriment of content, consistency, and argument. More attention is given to the way an economic theory or hypothesis is presented, while its contents are neglected (Blaug, 1998). In addition, the results of this mathematical formalization program have been at best insufficient or inadequate. It is commonly believed that when designing or using an economic-mathematical model, economic theory is being made, but it is often forgotten that most of the relations proposed in a model are relationships of a mathematical nature, albeit with an economic content or meaning. All inferences are obtained mathematically, and little attention is given to whether these variables, concepts and functional relations have some resemblance or correlation with an observation of the economic world. By failing to address this fundamental fact, economists have failed to emulate the methods of natural sciences in a crucial aspect, which is to consider the correct use of the dimensions. This leads to a definitive conclusion that is necessary to highlight in order to appropriately judge the current role of economic modelling: certain parts of the economic theory may contain important analytical flaws, and such defects are transmitted to the models based on this theory.

This paper has shown how Dimensional Analysis can be applied to models of economic growth; through the dimensional analysis of the Neoclassical model, a new system of differential equations was obtained from the corrected model, Eqs. (31-32), which lacks the dimensional inconsistency of the original model. However, a price has been paid with respect to simplicity, since now the equations that describe the dynamics of the optimal trajectories of consumption and accumulation of capital have become more complex, given the incorporation of the population growth rate $n$ within the production and utility functions.

The incorporation of the population growth rate does not preclude an immediate analysis of the equilibrium dynamics of the model; however, if this growth rate ceased to be constant and depended on other factors, the analytical complexity of the model would increase significantly and would prevent the use of classical methods of stability analysis.[28]

Nevertheless these implications, the dynamics presented by the new system of equidimensional equations of the model may be richer and more complex in relation to the behaviour of the accumulation of physical capital per capita, because now this accumulation will depend directly on the dynamics of the population growth rate, $n$, which would introduce a strong non-linearity in the dynamics of the model. As previously mentioned, a thorough analysis of the dynamical properties of the corrected model is beyond the scope of this article and will be the subject of future research.

Lastly, the possible lines of research that can be developed around the topics discussed in this work are indicated. It is considered necessary to analyse in depth the microeconomic and macroeconomic fundamentals of the neoclassical economic theory to identify whether the omission of Dimensional Analysis is an isolated event or occurs systematically in the theory. In addition, the principles of Dimensional Analysis should be studied more profoundly in order to direct these to the field of Economics, as there are still several theoretical questions to be answered at this point.

---

[28] Unlike the original Neoclassical model, in which an analysis of local or global stability can be made with relative ease. See Acemoglu (2009) or Barro & Sala-i-Martin (2009).